\documentclass[12pt]{article}
\usepackage{latexsym}
\usepackage{amsmath}
\usepackage{epsfig}
\title{Graviton Physics}
\author{Barry R. Holstein\\
Department of Physics---LGRT\\
University of Massachusetts\\
Amherst, MA  01003}

\begin{document}
\begin{titlepage}
\maketitle
\begin{abstract}
The interactions of gravitons with matter are calculated in
parallel with the familiar photon case.  It is shown that graviton
scattering amplitudes can be factorized into a product of familiar
electromagnetic forms, and cross sections for various reactions
are straightforwardly evaluated using helicity methods.
\end{abstract}
\end{titlepage}
\section{Introduction}
The calculation of photon interactions with matter is a staple in
an introductory (or advanced) quantum mechanics course.  Indeed
the evaluation of the Compton scattering cross section is a
standard exercise in relativistic quantum mechanics, since gauge
invariance together with the masslessness of the photon allow the
results to be presented in terms of simple analytic
forms\cite{hol}.

On the surface, a similar analysis should be applicable to the
interactions of gravitons.  Indeed, like photons, such particles
are massless and subject to a gauge invariance, so that similar
analytic results for graviton cross sections can be expected.
Also, just as virtual photon exchange leads to a detailed
understanding of electromagnetic interactions between charged
systems, a careful treatment of virtual graviton exchange allows
an understanding not just of Newtonian gravity, but also of
spin-dependent phenomena associated with general relativity which
are to be tested in the recently launched gravity probe
B\cite{hol1}.  However, despite this obvious parallel, examination
of quantum mechanics texts reveals that (with one
exception\cite{ari}) the case of graviton interactions is not
discussed in any detail. There are at least three reasons for this
situation:
\begin{itemize}
\item [i)] the graviton is a spin-two particle, as opposed to the
spin-one photon, so that the interaction forms are somewhat more
complex, involving symmetric and traceless second rank tensors
rather than simple Lorentz four-vectors;
\item [ii)] there exist few experimental results with which to compare
the theoretical calculations;
\item [iii)] as we will see later in some processes,
in order to guarantee gauge invariance one must include, in
addition to the usual Born and seagull diagrams, the contribution
from a graviton pole term, involving a triple-graviton coupling.
This vertex is a sixth rank tensor and contains a multitude of
kinematic forms.
\end{itemize}
However, recently, using powerful (string-based) techniques, which
simplify conventional quantum field theory calculations, it has
recently been demonstrated that the elastic scattering of
gravitons from an elementary target of arbitrary spin must
factorize\cite{fac}, a feature that had been noted ten years
previously by Choi et al. based on gauge theory
arguments\cite{kor}. This factorization permits a relatively
painless evaluation of the various graviton amplitudes. Below we
show how this factorization comes about and we evaluate some
relevant cross sections.  Such calculations can be used as an
interesting auxiliary topic within an advanced quantum mechanics
course

In the next section we review the simple electromagnetic case and
develop the corresponding gravitational formalism.  In section 3
we give the factorization results and calculate the relevant cross
sections, and our results are summarized in a concluding section
4.  Two appendices contain some of the formalism and calculational
details.

\section{Photon Interactions: a Lightning Review}

Before treating the case of gravitons it is useful to review the
case of photon interactions, since this familiar formalism can be
used as a bridge to our understanding of the gravitational case.
We begin by generating the photon interaction Lagrangian, which is
accomplished by writing down the free matter Lagrangian together
with the minimal substitution\cite{jdj}
$$i\partial_\mu\longrightarrow iD_\mu\equiv i\partial_\mu-eA_\mu$$
where $e$ is the particle charge and $A_\mu$ is the photon field.
As examples, we discuss below the case of a scalar field and a
spin 1/2 field, since these are familiar to most readers. Thus,
for example, the Lagrangian for a free charged Klein-Gordon field
is known to be
\begin{equation}
{\cal
L}=\partial_\mu\phi^\dagger\partial^\mu\phi+m^2\phi^\dagger\phi
\end{equation}
which becomes
\begin{equation}
{\cal L}=(\partial_\mu
-ieA_\mu)\phi^\dagger(\partial^\mu+ieA^\mu)\phi+m^2\phi^\dagger\phi
\end{equation}
after the minimal substitution.  The corresponding interaction
Lagrangian can then be identified---
\begin{equation}
{\cal
L}_{int}=-ieA_\mu(\partial^\mu\phi^\dagger\phi-\phi^\dagger\partial^\mu\phi)+e^2A^\mu
A_\mu\phi^\dagger\phi
\end{equation}
Similarly, for spin 1/2, the free Dirac Lagrangian
\begin{equation}
{\cal L}=\bar{\psi}(i\not\!{\nabla}-m)\psi
\end{equation}
becomes
\begin{equation}
{\cal L}=\bar{\psi}(i\not\!{\nabla}-e\not\!\!{A}-m)\psi
\end{equation}
whereby the interaction Lagrangian is found to be
\begin{equation}
{\cal L}=-e\bar{\psi}\not\!\!{A}\psi
\end{equation}
The single-photon vertices are then
\begin{equation}
<p_f|V_{em}^\mu|p_i>_{S=0}=e(p_f+p_i)^\mu
\end{equation}
for spin zero and
\begin{equation}
<p_f|V_{em}^\mu|p_i>_{S={1\over 2}}=e\bar{u}(p_f)\gamma^\mu u(p_i)
\end{equation}
for spin 1/2---and the amplitudes for photon (Compton)
scattering---{\it cf}. Figure 1---can be calculated.  In the case
of spin zero, {\it three} diagrams are involved---two Born terms
and a seagull---and the total amplitude is
\begin{equation}
{\rm Amp}_{Compton}(S=0)=2e^2\left[{2\epsilon_{i}\cdot
p_i\epsilon_f^*\cdot p_f\over p_i\cdot k_i}-{\epsilon_i\cdot
p_f\epsilon_f^*\cdot p_i\over p_i\cdot
k_f}-\epsilon_f^*\cdot\epsilon_i\right]
\end{equation}

\begin{figure}
\begin{center}
\epsfig{file=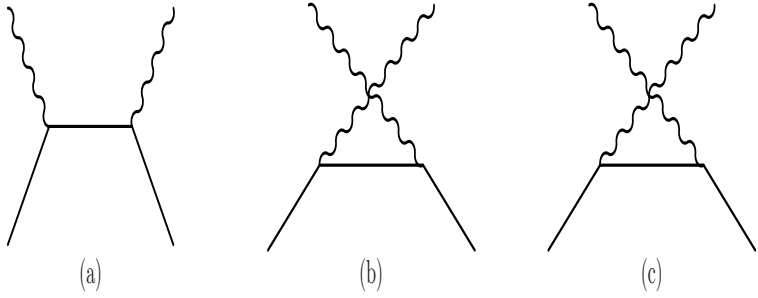,height=4cm,width=10cm} \caption{Diagrams
relevant to Compton scattering.}
\end{center}
\end{figure}

Note that all three diagrams must be included in order to satisfy
the stricture of gauge invariance, which requires that the
amplitude be unchanged under a gauge change
$$\epsilon_\mu\longrightarrow \epsilon_\mu+\lambda k_\mu$$
Indeed, we easily verify that under such a change for the incident
photon
\begin{eqnarray}
\delta {\rm Amp}_{Compton}(S=0)&=&\lambda 2e^2\left[{p_i\cdot
k_i\epsilon_f^*\cdot p_f\over k_i\cdot p_i}-{k_i\cdot
p_f\epsilon_f\cdot p_i\over p_f\cdot
k_i}-k_i\cdot\epsilon_f^*\right]\nonumber\\
&=&\lambda 2e^2\epsilon_f^*\cdot(p_f-p_i-k_i)=\lambda
2e^2\epsilon_f^*\cdot k_f=0
\end{eqnarray}
In the case of spin 1/2, there exists no seagull diagram and only
the two Born diagrams exist, yielding\cite{bj}
\begin{equation}
{\rm Amp}_{Compton}(S={1\over 2})=e^2\bar{u}(p_f)
\left[{\not\!{\epsilon}_f^*(\not\!{p}_i+\not\!{k}_i+m)\not\!{\epsilon}_i\over
2p_i\cdot
k_i}-{\not\!{\epsilon}_i(\not\!{p}_f-\not\!{k}_i+m)\not\!{\epsilon}_f^*\over
2p_f\cdot k_i}\right]u(p_i)
\end{equation}
Again, one can easily verify that this amplitude is
gauge-invariant---
\begin{eqnarray}
\delta{\rm Amp}_{Compton}(S={1\over 2})&=&\lambda
e^2\bar{u}(p_f)\left[{\not\!{\epsilon_f}^*(\not\!{p}_i+\not\!{k}_i+m)\not\!{k}_i\over
2p_i\cdot
k_i}-{\not\!{k}_i(\not\!{p}_f-\not\!{k}_i+m)\not\!{\epsilon}^*_f\over
2p_f\cdot k_i }\right]u(p_i)\nonumber\\
&=&\bar{u}(p_f)\left[\not\!{\epsilon}_f^*-\not\!{\epsilon}_f^*\right]u(p_i)=0
\end{eqnarray}

The corresponding cross sections can then be found via standard
methods, as shown in many texts\cite{bj}.  The results are usually
presented in the laboratory frame---$p_i=(m,\vec{0})$---wherein
the incident and final photon energies are related by
\begin{equation}
\omega_f={\omega_i\over 1+2{\omega_i\over m}\sin^2{1\over
2}\theta}
\end{equation}
where $\theta$ is the scattering angle.  For unpolarized
scattering, we sum (average) over final (initial) spins via
\begin{equation}
\sum_\lambda
\epsilon_\lambda^{*\mu}\epsilon_\lambda^\nu=-\eta^{\mu\nu},\quad
\sum_s u(p,s)_i\bar{u}(p,s)_j={(\not\!{p}+m)_{ij}\over 2m}
\end{equation}
and the resulting cross sections are well known---
\begin{equation}
{d\sigma_{lab}(S=0)\over d\Omega}={\alpha^2\over
m^2}({\omega_f\over \omega_i})^2{1\over
2}(1+\cos^2\theta)\label{eq:s0}
\end{equation}
and
\begin{eqnarray}
{d\sigma_{lab}(S=1/2)\over d\Omega}&=&{\alpha\over
2m^2}({\omega_f\over \omega_i})^2[{\omega_f\over
\omega_i}+{\omega_i\over \omega_f}-1+\cos^2\theta)]\nonumber\\
&=&{\alpha^2\over m^2}{\omega_f^2\over \omega_i^2}[{1\over 2
}(1+\cos^2\theta)(1+2{\omega_i\over m}\sin^2{1\over
2}\theta)+2{\omega_i^2\over m^2}\sin^4{1\over 2}\theta]\nonumber\\
\quad\label{eq:s1}
\end{eqnarray}\label{eq:std}

\subsection{Helicity Methods}

For use in the gravitational case it is useful to derive these
results in an alternative fashion, using the so-called "helicity
formalism,"  wherein one decomposes the amplitude in terms of
components of definite helicity\cite{hel}.  Here helicity is
defined by the projection of the particle spin along its momentum
direction. In the case of a photon moving along the z-direction,
we choose states
\begin{equation}
\epsilon_i^{\lambda_i}=-{\lambda_i\over
\sqrt{2}}(\hat{x}+i\lambda_i \hat{y}),\qquad \lambda_i=\pm
\end{equation}
while for a photon moving in the direction
$$\hat{k}_f=\sin\theta\hat{x}+\cos\theta\hat{z}$$
we use states
\begin{equation}
\epsilon_f^{\lambda_f}=-{\lambda_f\over
\sqrt{2}}(\cos\theta\hat{x}+
i\lambda_f\hat{y}-\sin\theta\hat{z}),\quad \lambda_f=\pm
\end{equation}
Working in the center of mass frame we can then calculate the
amplitude for transitions between states of definite helicity.
Using
$$\epsilon_i^\pm\cdot p_f=-\epsilon_f^{*\pm}\cdot p_i=\mp {p\over \sqrt{2}}\sin\theta$$
$$\epsilon_f^{*\pm}\cdot\epsilon_i^{\pm}=-{1\over 2}(1+\cos\theta),\quad
\epsilon_f^{*\pm}\cdot\epsilon_i^{\mp}=-{1\over 2}(1-\cos\theta)$$
we find for spin zero Compton scattering
\begin{eqnarray}
A_{++}&=&A_{--}=-e^2\left(1+\cos\theta+{p^2\sin^2\theta\over p_i\cdot k_f}\right)\nonumber\\
A_{+-}&=&A_{-+}=-e^2\left(1-\cos\theta-{p^2\sin^2\theta\over
p_i\cdot k_f}\right)
\end{eqnarray}
While these results can be found by direct calculation, the
process can be simplified by realizing that under a parity
transformation the momentum reverses but the spin stays the same.
Thus the helicity reverses, so parity conservation assures the
equality of $A_{a,b}$ and $A_{-a,-b}$ while under time reversal
both spin and momentum change sign, as do initial and final
states, guaranteeing that helicity amplitudes are
symmetric---$A_{a,b}=A_{b,a}$.

Using the standard definitions
$$s=(p_i+k_i)^2,\quad t=(k_i-k_f)^2,\quad u=(p_i-k_f)^2$$
it is easy to see from simple kinematical considerations that
\begin{eqnarray}
p^2={(s-m^2)^2\over 4s},\quad \cos{1\over
2}\theta&=&{((s-m^2)^2+st)^{1\over 2}\over
 s-m^2}={(m^4-su)^{1\over 2}\over s-m^2}\nonumber\\
  \sin{1\over 2}\theta&=&{(-st)^{1\over 2}\over (s-m^2)}
 \end{eqnarray}
We can write then\cite{jac}
\begin{eqnarray}
A_{++}&=&A_{--}=2e^2{(s-m^2)^2+st\over (s-m^2)(u-m^2)}\nonumber\\
A_{+-}&=&A_{-+}=2e^2{-m^2t\over (s-m^2)(u-m^2)}
\end{eqnarray}
(It is interesting that the general form of these amplitudes
follows from simple kinematical constraints, as shown in ref.
\cite{van}.) The cross section can now be written in terms of
Lorentz invariants as
\begin{eqnarray}
{d\sigma\over dt}&=&{1\over 16\pi(s-m^2)^2}{1\over
2}\sum_{i,j=\pm}|A_{ij}|^2\nonumber\\
&=&4e^4{(m^4-su)^2+m^4t^2\over 16\pi (s-m^2)^4(u-m^2)^2}
\end{eqnarray}
and can be evaluated in any desired frame.  In particular, in the
laboratory frame we have
\begin{eqnarray}
s-m^2&=&2m\omega_i,\quad u-m^2=-2m\omega_f\nonumber\\
m^4-su&=&4m^2\omega_i\omega_f\cos^2{1\over 2}\theta,\quad
m^2t=-4m^2\omega_i\omega_f\sin^2{1\over 2}\theta
\end{eqnarray}
Since
\begin{equation}
{dt\over d\Omega}={d\over 2\pi
d\cos\theta}\left(-{2\omega_i^2(1-\cos\theta)\over
1+{\omega_i\over m}(1-\cos\theta)}\right)={\omega_f^2\over \pi}
\end{equation}
the laboratory cross section is found to have the form
\begin{equation}
{d\sigma_{lab}\over d\Omega}(S=0)={d\sigma(S=0)\over dt}{dt\over
d\Omega}={\alpha^2\over m^2}{\omega_f^2\over
\omega_i^2}(\cos^4{1\over 2}\theta+\sin^4{1\over
2}\theta)\label{eq:sp}
\end{equation}
and, using the identity
$$\cos^4{1\over 2}\theta+\sin^4{1\over 2}\theta={1\over 2}(1+\cos^2\theta)$$
Eq. \ref{eq:sp} is seen to be identical to Eq. \ref{eq:s0} derived
by conventional means.

A corresponding analysis can be performed for spin 1/2. Working
again in the center of mass frame and using helicity states for
both photons and spinors, one can calculate the various
amplitudes. In this case it is convenient to define the photon as
the "target" particle, so that the corresponding polarization
vectors are
\begin{eqnarray}
\epsilon_i^{\lambda_i}&=&-{\lambda_i\over
\sqrt{2}}(-\hat{x}+i\lambda_i\hat{y})\nonumber\\
\epsilon_f^{\lambda_f*}&=&-{\lambda_f\over
\sqrt{2}}(\sin\theta\hat{x}+\cos\theta\hat{z}-i\lambda_f\hat{y})
\end{eqnarray}
for initial (final) state helicity $\lambda_i$ ($\lambda_f$).
Working in the center of mass, the corresponding helicity
amplitudes can then be evaluated via
\begin{equation}
B_{s_f\lambda_f;s_i\lambda_i}=\bar{u}(p_f,s_f)[{2\epsilon_i\cdot
p_i\not\!{\epsilon}_f^*\over p_i\cdot k_i}-{2\epsilon_i\cdot
p_f\not\!{\epsilon}_f^*\over p_i\cdot
k_f}+{\not\!{\epsilon}_f^*\not\!{k}_i\not\!{\epsilon}_i\over
p_i\cdot
k_i}-{\not\!{\epsilon}_i\not\!{k}_i\not\!{\epsilon}_f^*\over
p_i\cdot k_f}]u(p_i,s_i)
\end{equation}
Useful identities in this evaluation are
\begin{eqnarray}
{\cal O
}_1&\equiv&\not\!{\epsilon}_f^*\not\!{k}_i\not\!{\epsilon}_i={p\lambda_i\lambda_f\over
2}\left(\begin{array}{cc} A+\Sigma&\lambda_i(A+\Sigma)\\
-\lambda_i(A+\Sigma)&-(A+\Sigma)
\end{array}\right)\nonumber\\
{\cal O
}_2&\equiv&\not\!{\epsilon}_i\not\!{k}_i\not\!{\epsilon}_f^*={p\lambda_i\lambda_f\over
2}\left(\begin{array}{cc} A-\Sigma&-\lambda_i(A-\Sigma)\\
\lambda_i(A-\Sigma)&-(A-\Sigma)
\end{array}\right)\nonumber\\
{\cal O}_3&\equiv&\epsilon_i\cdot
p_f\not\!{\epsilon}_f^*={p\sin\theta\lambda_i\lambda_f\over 2}
\left(\begin{array}{cc} 0&\Upsilon\\-\Upsilon&0\end{array}\right)
\end{eqnarray}
where
\begin{eqnarray}
A&=&\lambda_i\lambda_f+\cos\theta\nonumber\\
\Sigma&=&(\cos\theta\lambda_i+\lambda_f)\sigma_z+\lambda_i\sin\theta\sigma_x
-i\sin\theta\sigma_y\nonumber\\
\Upsilon&=&-\cos\theta\sigma_x+\sin\theta\sigma_z-i\lambda_f\sigma_y
\end{eqnarray}
We have then
\begin{eqnarray}
B_{s_f\lambda_f;s_i\lambda_i}&=&{E+m\over
2m}\left(\begin{array}{cc}\chi_f^\dagger&{-2s_fp\over
E+m}\chi_f^\dagger\end{array}\right)\nonumber\\
&\times&\left[{1\over s-m^2}{\cal O}_1-{1\over u-m^2}({\cal
O}_2+{\cal O}_3)\right]\left(\begin{array}{c}\chi_i\\{2s_ip\over
E+m} \chi_i\end{array}\right)
\end{eqnarray}
and, after a straightforward (but tedious) exercise, one finds the
amplitudes\cite{hen}
\begin{eqnarray}
B_{{1\over 2}1;{1\over 2}1}&=&B_{-{1\over 2}-1;-{1\over 2}-1}
={2\sqrt{s}e^2p\cos{1\over 2}\theta\over m^2-u}(-{1\over
m}+{m\over s}\sin^2{1\over 2}\theta)\nonumber\\
B_{{1\over 2}1;{1\over 2}-1}&=&B_{-{1\over 2}1;-{1\over 2}-1}=
B_{{1\over 2}-1;{1\over 2}1}=B_{-{1\over 2}-1;-{1\over 2}1}
=-{2e^2mp\over (m^2-u)\sqrt{s}}\sin^2{1\over 2}\theta\cos{1\over
2}\theta\nonumber\\
B_{{1\over 2}-1;{1\over 2}-1}&=&B_{-{1\over 2}1;-{1\over 2}1}
={-2\sqrt{s}e^2p\over m(m^2-u)}\cos^3{1\over 2}\theta\nonumber\\
B_{{1\over 2}1;-{1\over 2}1}&=&-B_{-{1\over 2}-1;{1\over 2}-1}=
B_{-{1\over 2}1;{1\over 2}1}=-B_{{1\over 2}-1;-{1\over 2}-1}
=-{2e^2p\over (m^2-u)}\sin{1\over 2}\theta\cos^2{1\over
2}\theta\nonumber\\
B_{{1\over 2}-1;-{1\over 2}1}&=&-B_{-{1\over 2}1;{1\over 2}-1}
={-2e^2p\over m^2-u}\sin^3{1\over 2}\theta\nonumber\\
B_{{1\over 2}1;-{1\over 2}-1}&=&-B_{-{1\over 2}-1;{1\over 2}1}
=-{2e^2m^2p\over s(m^2-u)}\sin^3{1\over 2}\theta
\end{eqnarray}
Summing (averaging) over final (initial) spin 1/2 states we define
spin-averaged photon helicity quantities
\begin{eqnarray}
|B_{++}|^2_{av}&=&|B_{--}|^2_{av}={2p^2e^4s\sin^2{1\over
2}\theta\over m^2(m^2-u)^2}\left[1+\cos^4{1\over
2}\theta+\sin^4{1\over
2}\theta{m^4\over s^2}(1-2{s\over m^2})\right]\nonumber\\
|B_{+-}|^2_{av}&=&|B_{-+}|^2_{av}={2p^2e^4\sin^4{1\over
2}\theta\over (m^2-u)^2}\left[2{m^2\over s}\cos^2{1\over
2}\theta+(1+{m^4\over s^2})\sin^2{1\over 2}\theta\right]
\end{eqnarray}
which, in terms of invariants, have the form
\begin{eqnarray}
|B_{++}|^2_{av}&=&|B_{--}|^2_{av}={e^4\over
2m^2(u-m^2)^2(s-m^2)^2}(m^4-su)(2(m^4-su)+t^2)\nonumber\\
|B_{+-}|^2_{av}&=&|B_{-+}|^2_{av}={e^4t^2(2m^2-t)\over
2(s-m^2)^2(u-m^2)^2}
\end{eqnarray}
The laboratory frame cross section can be determined as before
(note that this expression differs from Eq. \ref{eq:sp} by the
factor $4m^2$, which is due to the different normalizations for
fermion and boson states---$m/E$ for fermions and $1/2E$ for
bosons)
\begin{eqnarray}
{d\sigma(S=1/2)\over d\Omega}&=&{4m^2\omega_f^2\over
16\pi^2(s-m^2)^2}[|B_{++}|^2_{av}+|B_{+-}|^2_{av}]\nonumber\\
&=&{\alpha^2\over m^2}{\omega_f^2\over \omega_i^2}[{1\over 2
}(1+\cos^2\theta)(1+2{\omega_i\over m}\sin^2{1\over
2}\theta)+2{\omega_i^2\over m^2}\sin^4{1\over 2}\theta]\nonumber\\
\quad
\end{eqnarray}
which is seen to be identical to the previously form---Eq.
\ref{eq:s1}.

So far, all we have done is to derive the usual forms for Compton
cross sections by non-traditional means.  However, in the next
section we shall see how the use of helicity methods allows the
derivation of the corresponding graviton cross sections in an
equally straightforward fashion.

\section{Gravitation}

The theory of graviton interactions can be developed in direct
analogy to that of electromagnetism.  Some of the details are
given in Appendix A.  Here we shall be content with a brief
outline.

Just as the electromagnetic interaction can be written in terms of
the coupling of a vector current $j_\mu$ to the vector potential
$A^\mu$ with a coupling constant given by the charge $e$
\begin{equation}
{\cal L}_{int}=-ej_\mu a^\mu
\end{equation}
the gravitational interaction can be described in terms of the
coupling of the energy-momentum tensor $T_{\mu\nu}$ to the
gravitational field $h^{\mu\nu}$ with a coupling constant $\kappa$
\begin{equation}
{\cal L}_{int}=-{1\over 2}\kappa T_{\mu\nu}h^{\mu\nu}
\end{equation}
Here the field tensor is defined in terms of the metric via
\begin{equation}
g_{\mu\nu}=\eta_{\mu\nu}+\kappa h_{\mu\nu}
\end{equation}
while $\kappa$ is defined in terms of Newton's constant via
$\kappa^2=32\pi G$.  The energy-momentum tensor is defined in
terms of the free matter Lagrangian via
\begin{equation}
T_{\mu\nu}={2\over \sqrt{-g}}{\delta {\sqrt{-g}\cal L}_{int}\over
\delta g^{\mu\nu}}
\end{equation}
where
\begin{equation}
\sqrt{-g}=\sqrt{-{\rm det}g}=\exp{1\over 2}{\rm tr}{\rm log} g
\end{equation}
is the square root of the determinant of the metric. This
prescription yields the forms
\begin{equation}
T_{\mu\nu}=\partial_\mu\phi^\dagger\partial_\nu\phi+\partial_\nu\phi^\dagger
\partial_\mu\phi-g_{\mu\nu}(\partial_\mu\phi^\dagger\partial^\mu\phi-m^2\phi^\dagger\phi)
\end{equation}
for a scalar field and
\begin{equation}
T_{\mu\nu}=\bar{\psi}[{1\over 4}\gamma_\mu
i\overleftrightarrow{\nabla}_\nu+{1\over 4}\gamma_\nu
i\overleftrightarrow{\nabla}_\mu-g_{\mu\nu}({i\over
2}\not\!{\overleftrightarrow{\nabla}}-m)]\psi
\end{equation}
for spin 1/2, where we have defined
\begin{equation}
\bar{\psi}i\overleftrightarrow{\nabla}_\mu\psi\equiv
\bar{\psi}i\nabla_\mu\psi-(i\nabla_\mu\bar{\psi})\psi
\end{equation}
The matrix elements of $T_{\mu\nu}$ can now be read off as
\begin{equation}
<p_f|T_{\mu\nu}|p_i>_{S=0}=p_{f\mu}p_{i\nu}+p_{f\nu}p_{i\mu}-\eta_{\mu\nu}(p_f\cdot
p_i-m^2)
\end{equation}
and
\begin{equation}
<p_f|T_{\mu\nu}|p_i>_{S={1\over 2 }}=\bar{u}(p_f)[{1\over 4
}\gamma_\mu(p_f+p_i)_\mu+{1\over 4}\gamma_\nu(p_f+p_i)_\mu]u(p_i)
\end{equation}
We shall work in harmonic (deDonder) gauge which satisfies, in
lowest order,
\begin{equation}
\partial^\mu h_{\mu\nu}={1\over 2}\partial_\nu h\label{eq:dd}
\end{equation}
where
\begin{equation}
h={\rm tr}h_{\mu\nu}
\end{equation}
 and yields a graviton propagator
\begin{equation}
D_{\alpha\beta;\gamma\delta}(q)={i\over
2q^2}(\eta_{\alpha\gamma}\eta_{\beta\delta}+\eta_{\alpha\delta}\eta_{\beta\gamma}
-\eta_{\alpha\beta}\eta_{\gamma\delta})
\end{equation}
Then just as the (massless) photon is described in terms of a
spin-one polarization vector $\epsilon_\mu$ which can have
projection (helicity) either plus or minus one along the momentum
direction, the (massless) graviton is a spin two particle which
can have the projection (helicity) either plus or minus two along
the momentum direction.  Since $h_{\mu\nu}$ is a symmetric tensor,
it can be described in terms of a simple product of unit spin
polarization vectors---
\begin{eqnarray}
{\rm helicity}&=&+2:\quad
h^{(2)}_{\mu\nu}=\epsilon^{+}_\mu \epsilon^{+}_\nu\nonumber\\
{\rm helicity}&=&-2:\quad h^{(-2)}_{\mu\nu}=\epsilon^{-}_\mu
\epsilon^{-}_\nu\label{eq:he}
\end{eqnarray}
and just as in electromagnetism, there is a gauge condition---in
this case Eq. \ref{eq:dd}---which must be satisfied. Note that the
helicity states given in Eq. \ref{eq:he} are consistent with the
gauge requirement, since
\begin{equation}
\eta^{\mu\nu}\epsilon_\mu\epsilon_\nu=0,\quad{\rm and} \quad
k^\mu\epsilon_\mu=0
\end{equation}
With this background we can now examine various interesting
reactions involving gravitons, as we detail below.

\subsection{Graviton Photoproduction}

Before dealing with our ultimate goal, which is the treatment of
graviton Compton scattering, we first warm up with a simpler
process---that of graviton photoproduction
$$\gamma+s\rightarrow g+s,\quad \gamma+f\rightarrow g+f$$
The relevant diagrams are shown in Figure 2 and include the Born
diagrams accompanied by a seagull and by the photon pole.  The
existence of a seagull is required by the feature that the
energy-momentum tensor is momentum-dependent and therefore yields
a contact interactions when the minimal substitution is made,
yielding the amplitudes
\begin{equation}
<p_f;k_f,\epsilon_f\epsilon_f|T|p_i;k_i,\epsilon_i>_{seagull}=\kappa
e\left\{
\begin{array}{cc}
-2\epsilon_f^*\cdot\epsilon_i \epsilon_f^*\cdot(p_f+p_i)&S=0\\
\bar{u}(p_f)\not\!{\epsilon}_f^*\epsilon_f^*\cdot\epsilon_iu(p_i)&
S={1\over 2}\end{array}\right.
\end{equation}
For the photon pole diagram we require a new ingredient, the
graviton-photon coupling, which can be found from the expression
for the photon energy-momentum tensor\cite{jdj}
\begin{equation}
T_{\mu\nu}=-F_{\mu\alpha}F^\alpha_\nu+{1\over
2}g_{\mu\nu}F_{\alpha\beta}F^{\alpha\beta}
\end{equation}
This yields the photon pole term
\begin{eqnarray}
<p_f;k_f,\epsilon_f\epsilon_f|T||p_i;k_i\epsilon_i>_{\gamma-pole}=
e<p_f|j_\alpha|p_i>{1\over (p_f-p_i)^2}\nonumber\\
\times {\kappa\over 2}[2\epsilon_f^{*\alpha}(k_f\cdot
k_i\epsilon_f^*\cdot\epsilon_i-\epsilon_f^*\cdot
k_i\epsilon_i\cdot k_f)+2\epsilon_i\cdot
k_f(\epsilon_f^*\cdot\epsilon_i k_f^\alpha-\epsilon_i\cdot
k_f\epsilon_f^{*\alpha})]
\end{eqnarray}

\begin{figure}
\begin{center}
\epsfig{file=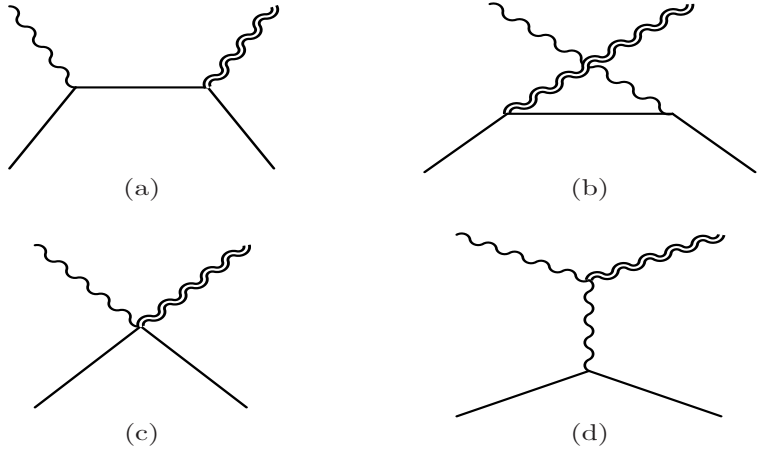,height=6cm,width=10cm} \caption{Diagrams
relevant to graviton photoproduction. }
\end{center}
\end{figure}

Adding the four diagrams together, we find (after considerable but
simple algebra---{\it cf.} Appendix B) a remarkably simple result
\begin{equation}
<p_f;k_f,\epsilon_f\epsilon_f|T|p_i;k_i,\epsilon_i>= H\times
\left(\epsilon_{f\alpha}^*\epsilon_{i\beta}
T_{Compton}^{\alpha\beta}(S)\right)\label{eq:gi}
\end{equation}
where $H$ is the factor
\begin{equation}
H={\kappa\over 4e}{\epsilon_f^*\cdot p_fk_f\cdot
p_i-\epsilon_f^*\cdot p_ik_f\cdot p_f\over k_i\cdot k_f }
\end{equation}
and $\epsilon_{f\alpha}^*\epsilon_{i\beta}
T_{Compton}^{\alpha\beta}(S)$ is the Compton scattering amplitude
for particles of spin S calculated in the previous section.  The
gauge invariance of Eq. \ref{eq:gi} is obvious, since it follows
directly from the gauge invariance already shown for the
corresponding photon amplitudes together with that of the factor
$H$ under
$$\epsilon_f\rightarrow \epsilon_f+\lambda k_f.$$
Also we note that in Eq. \ref{eq:gi} the factorization condition
mentioned in the introduction is made manifest, and consequently
the corresponding cross sections can be obtained trivially. In
principle, one can use conventional techniques, but this is
somewhat challenging in view of the tensor structure of the
graviton polarization vector. However, factorization means that
helicity amplitudes for graviton photoproduction are simple
products of the corresponding photon amplitudes times the
universal factor $H$, and the cross sections are then given by the
simple photon forms times the universal factor $H^2$.  In the CM
frame we have $\epsilon_f^*\cdot p_i=-\epsilon_f^*\cdot k_i$ and
the factor $H$ assumes the form
\begin{equation}
|H|={\kappa\over 4e}|{\epsilon_f^*\cdot k_i k_f\cdot p_f\over
k_i\cdot k_f}|={\kappa\over 4e}{p\sin\theta\over \sqrt {2}}
{s-m^2\over -t}={\kappa\over 2e}\left[{m^4-st\over
-2t}\right]^{1\over 2}
\end{equation}
In the lab frame this takes the form
\begin{equation}
|H_{lab}|^2={\kappa^2m^2\over 8e^2}{\cos^2{1\over 2}\theta\over
\sin^2{1\over 2}\theta}
\end{equation}
and the graviton photoproduction cross sections are found to be
\begin{eqnarray}
&&{d\sigma\over d\Omega}=G\alpha\cos^2{1\over
2}\theta\nonumber\\
&\times& \left\{\begin{array}{ll} ({\omega_f\over
\omega_i})^2[{\rm ctn}^2{1\over 2}\theta\cos^2{1\over
2}\theta+\sin^2{1\over
2}\theta]& S=0\\
({\omega_f\over \omega_i})^3[({\rm ctn}^2{1\over
2}\theta\cos^2{1\over 2}\theta+\sin^2{1\over
2}\theta)+{2\omega_i\over m}(\cos^4{1\over 2}\theta+\sin^4{1\over
2 }\theta)+2{\omega_i^2\over m^2}\sin^2{1\over 2}\theta]&S=1/2
\end{array}\right.\nonumber\\
\qquad
\end{eqnarray}
The form of the latter has previously been given by
Voronov.\cite{vor}

\subsection{Graviton Compton Scattering}

Finally, we can proceed to our primary goal, which is the
calculation of graviton Compton scattering.  In order to produce a
gauge invariant scattering amplitude in this case we require {\it
four} separate contributions, as shown in Figure 3.  Two of these
diagrams are Born terms and can be written down straightforwardly.
However, there are also seagull terms both for spin 0 and for spin
1/2, whose forms can be found in Appendix A.  For the scalar case
we have
\begin{eqnarray}
<p_f;k_f,\epsilon_f\epsilon_f|T|p_i;k_i\epsilon_i\epsilon_i>_{seagull}&=&\left({\kappa\over
2}\right)^2\left[-2 \epsilon_f^*\cdot \epsilon_i(\epsilon_i\cdot
p_i\epsilon_f^*\cdot p_f+\epsilon_i\cdot p_f\epsilon_f^*\cdot
p_i)\right.\nonumber\\&-&\left.{1\over
2}(\epsilon_f^*\cdot\epsilon_i)^2k_i\cdot k_f\right]
\end{eqnarray}
while in the case of spin 1/2
\begin{eqnarray}
<p_f;k_f,\epsilon_f\epsilon_f|T|p_i;k_i\epsilon_i\epsilon_i>_{seagull}&=&\left({\kappa\over
2}\right)^2\bar{u}(p_f)\nonumber\\
&\times&\left[{3\over
16}\epsilon_f^*\cdot\epsilon_i(\not\!{\epsilon_i}\epsilon_f^*\cdot(p_i+p_f)
+\not\!{\epsilon}_f^*\epsilon_i\cdot(p_i+p_f))\right.\nonumber\\
&+&\left.{i\over
16}\epsilon_f^*\cdot\epsilon_i\epsilon^{\rho\sigma\eta\lambda}
\gamma_\lambda\gamma_5(\epsilon_{i\eta}\epsilon_{f\sigma}^*k_{f\rho}-
\epsilon_{f\eta}\epsilon_{i\sigma}k_{i\rho})\right]u(p_i)\nonumber\\
\quad
\end{eqnarray}

\begin{figure}
\begin{center}
\epsfig{file=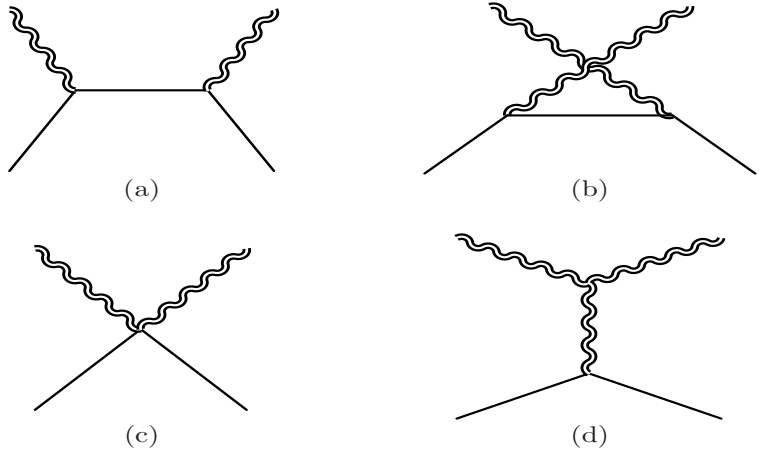,height=6cm,width=10cm} \caption{Diagrams
relevant for gravitational Compton scattering. }
\end{center}
\end{figure}

Despite the complex form of the various contributions, the final
form, which results (after considerable algebra-{\it cf}. Appendix
B) upon summation of the various components, is remarkably simple:
\begin{equation}
\epsilon_{f\alpha}\epsilon_{f\beta}M_{grav}^{\alpha\beta;\gamma\delta}
\epsilon_{i\gamma}\epsilon_{i\delta}=
F\times\left(\epsilon_{f\mu}^*\epsilon_{i\nu}
T^{\mu\nu}_{Compton}(S=0)\right)\times
\left(\epsilon_{f\alpha}^*\epsilon_{i\beta}
T^{\alpha\beta}_{Compton}(S)\right)\label{eq:fac}
\end{equation}
where $F$ is the universal factor
\begin{equation}
F={\kappa^2\over 8e^4}{p_i\cdot k_ip_i\cdot k_f\over k_i\cdot k_f}
\end{equation}
and the Compton amplitudes are those calculated in section 2.  The
gauge invariance of this form is again obvious from the
already-demonstrated gauge invariance of the photon amplitudes and
this is the factorized form guaranteed by general
arguments\cite{kor}. Again, it is in principle possible but very
challenging to evaluate the cross section by standard means, but
the result follows directly by the use of helicity methods.  From
the form of Eq. \ref{eq:fac} it is clear that the helicity
amplitudes for graviton scattering have the simple form of a
product of corresponding helicity amplitudes for spinless and spin
S Compton scattering. That is, for graviton scattering from a
spinless target we have
\begin{eqnarray}
|C_{++}|^2&=&|C_{--}|^2=F^2|A_{++}|^4\nonumber\\
|C_{+-}|^2&=&|C_{-+}|^2=F^2|A_{+-}|^4
\end{eqnarray}
while for scattering from a spin 1/2 target, we find for the
target-spin averaged helicity amplitudes
\begin{eqnarray}
|D_{++}|_{av}^2&=&|C_{--}|_{av}^2=F^2|A_{++}|^2|B_{++}|_{av}^2\nonumber\\
|C_{+-}|_{av}^2&=&|C_{-+}|_{av}^2=F^2|A_{+-}|^2|B_{+-}|_{av}^2
\end{eqnarray}
Here the factor F has the form
\begin{equation}
F={\kappa^2\over 8e^4}{(s-m^2)(u-m^2)\over t}
\end{equation}
whose laboratory frame value is
\begin{equation}
F_{lab}={\kappa^2m^2\over 8e^4}{1\over \sin^2{1\over 2}\theta}
\end{equation}
The corresponding laboratory cross sections are found then to be
\begin{eqnarray}
{d\sigma_{lab}\over d\Omega}(S=0)&=&{\omega_f^2\over
\pi}F^2{1\over
16\pi(s-m^2)^2}(|C_{++}|^2+|C_{+-}|^2)\nonumber\\
&=&G^2m^2({\omega_f\over \omega_i})^2[{\rm ctn}^4{1\over
2}\theta\cos^4{1\over 2}\theta+\sin^4{1\over 2}\theta]
\end{eqnarray}
for a spinless target and
\begin{eqnarray}
{d\sigma_{lab}\over d\Omega}(S={1\over 2})&=&{\omega_f^2\over
\pi}F^2{1\over
16\pi(s-m^2)^2}(|D_{++}|_{av}^2+|D_{+-}|_{av}^2)\nonumber\\
&=&G^2m^2({\omega_f\over \omega_i})^3[{\rm ctn}^4{1\over
2}\theta\cos^4{1\over 2}\theta+\sin^4{1\over
2}\theta)\nonumber\\
&+&2{\omega_i\over m}({\rm ctn}^2{1\over 2}\theta\cos^6{1\over
2}\theta+\sin^6{1\over 2}\theta)+2{\omega_i^2\over
m^2}(\cos^6{1\over 2}\theta+\sin^6{1\over 2}\theta)]\nonumber\\
\qquad
\end{eqnarray}
for a spin 1/2 target.  The latter form agrees with that given by
Voronov\cite{vor}.

We have given the results for unpolarized scattering from an
unpolarized target, but having the form of the helicity amplitudes
means that we can also produce cross sections involving polarized
photons or gravitons.  That is, however, a subject for a different
time and a different paper.

\section{Summary}

While the subject of photon interactions with charged particles is
a standard one in any quantum mechanics course, the same is not
true for that of graviton interactions with masses despite the
obvious parallels between these two topics.  The origin of this
disparity lies with the complications associated with the tensor
structure of gravity and the inherent nonlinearity of
gravitational theory.  We have argued above that this need not be
the case.  Indeed in an earlier work we showed how the parallel
between the exchange of {\it virtual} gravitons and photons could
be used in order to understand the phenomena of geodetic and
Lense-Thirring precession in terms of the the related spin-orbit
and spin-spin interactions in quantum electrodynamics\cite{hol1}.
In the present paper, we have shown how the treatment of graviton
scattering processes can benefit from use of this analogy.  Of
course, such amplitudes are inherently more complex, in that they
must involve tensor polarization vectors and the addition of
somewhat complex photon or graviton pole diagrams.  However, it is
remarkable that when all effects are added together, the resulting
amplitudes factorize into simple products of photon amplitudes
times kinematic factors.  Using helicity methods, this
factorization property then allows the relatively elementary
calculation of cross sections since they involve simple products
of the already known photon amplitudes times kinematical factors.
It is hoped that this remarkable result will allow introduction of
graviton reactions into the quantum mechanics cirriculum in at
least perhaps a special topics presentation.  In any case, the
simplicity associated with this result means that graviton
interactions can be considered a topic which is no longer only
associated with advanced research papers.

\begin{center}
{\bf Appendix A: Gravitational Formalism}
\end{center}
Here we present some of the basics of gravitational field theory.
Details can be found in various references\cite{wei,boh}.  The
full gravitational action is given by
\begin{equation}
S_g=\int d^4x \sqrt{-g}\left({1\over 16\pi G}R+{\cal
L}_m\right)\label{eq:ful}
\end{equation}
where ${\cal L}_m$ is the Lagrange density for matter and $R$ is
the scalar curvature.  Variation of Eq. \ref{eq:ful} via
$$g_{\mu\nu}\rightarrow \eta_{\mu\nu}+\kappa h_{\mu\nu}$$
yields the Einstein equation
\begin{equation}
R_{\mu\nu}-{1\over 2}g_{\mu\nu}R=-8\pi GT_{\mu\nu}
\end{equation}
where the energy-momentum tensor $T_{\mu\nu}$ is given by
\begin{equation}
T_{\mu\nu}={2\over \sqrt{-g}}{\partial\over \partial
g^{\mu\nu}}(\sqrt{-g}{\cal L}_m)
\end{equation}
We work in the weak field limit, with an expansion in powers of
the gravitational coupling $G$
\begin{eqnarray}
g_{\mu\nu}&\equiv&\eta_{\mu\nu}+
\kappa h_{\mu\nu}^{(1)}+\ldots\nonumber\\
g^{\mu\nu}&=&\eta^{\mu\nu}-\kappa h^{(1)\mu\nu}+\kappa^2
h^{(1)\mu\lambda} {h^{(1)}}_\lambda{}^\nu+\ldots
\end{eqnarray}
where here the superscript indicates the number of powers of $G$
which appear and indices are understood to be raised or lowered by
$\eta_{\mu\nu}$. We shall also need the determinant which is given
by
\begin{equation}
\sqrt{-g}=\exp{1\over 2}{\rm tr}\log\,g=1+{1\over 2}\kappa h^{(1)}
+\ldots
\end{equation}
The corresponding curvatures are given by
\begin{eqnarray}
R_{\mu\nu}^{(1)}&=&{\kappa\over 2}\left[\partial_\mu\partial_\nu
h^{(1)}+
\partial_\lambda\partial^\lambda
 h_{\mu\nu}^{(1)}-\partial_\mu\partial_\lambda
{h^{(1)}}^\lambda{}_\nu
-\partial_\nu\partial_\lambda {h^{(1)}}^\lambda{}_\mu\right]\nonumber\\
R^{(1)}&=&\eta^{\mu\nu}R^{(1)}_{\mu\nu}=\kappa\left[\Box
h^{(1)}-\partial_\mu\partial_\nu h^{(1)\mu\nu}\right]
\end{eqnarray}
In order to define the graviton propagator, we must make a gauge
choice and we shall work in harmonic (or deDonder)
gauge---$g^{\mu\nu} \Gamma^\lambda_{\mu\nu}=0$---which requires,
to first order in the field expansion,
\begin{equation}
0=\partial^\beta h^{(1)}_{\beta\alpha}-{1\over 2}\partial_\alpha
h^{(1)}\label{eq:gau}
\end{equation}
Using these results, the Einstein equation reads, in lowest order,
\begin{equation}
\Box h^{(1)}_{\mu\nu}-{1\over 2}\eta_{\mu\nu}\Box h^{(1)}
-\partial_\mu\left(\partial^\beta h^{(1)}_{\beta\nu}-{1\over
2}\partial_\nu h^{(1)}\right)-\partial_\nu\left(\partial^\beta
h^{(1)}_{\beta\mu}-{1\over 2}\partial_\mu h^{(1)}\right)=-16\pi
GT_{\mu\nu}^{\rm matt}
\end{equation}
which, using the gauge condition Eq. \ref{eq:gau}, can be written
as
\begin{equation}
\Box \left(h_{\mu\nu}^{(1)}-{1\over 2}\eta_{\mu\nu}h^{(1)}\right)
=-16\pi G T_{\mu\nu}^{\rm matt}
\end{equation}
or in the equivalent form
\begin{equation}
\Box h_{\mu\nu}^{(1)}=-16\pi G\left( T_{\mu\nu}^{\rm matt}-{1\over
2}\eta_{\mu\nu} T^{\rm matt}\right)
\end{equation}

\begin{center}
{\bf Gravitational Interactions: Spin 0}
\end{center}

The coupling to matter via one-graviton and two-graviton vertices
can be found by expanding the spin zero matter Lagrangian
\begin{equation}
\sqrt{-g}{\cal L}_m=\sqrt{-g}\left({1\over 2}D_\mu\phi
g^{\mu\nu}D_\nu\phi -{1\over 2}m^2\phi^2\right)
\end{equation}
via
\begin{eqnarray}
\sqrt{-g}{\cal L}_m^{(0)}&=&{1\over
2}(\partial_\mu\phi\partial^\mu\phi-m^2\phi^2)
\nonumber\\
\sqrt{-g}{\cal L}_m^{(1)}&=&-{\kappa\over
2}h^{(1)\mu\nu}\left(\partial_\mu\phi
\partial_\nu\phi-{1\over 2}\eta_{\mu\nu}(\partial_\alpha\phi
\partial^\alpha\phi-m^2\phi^2)\right)\nonumber\\
\sqrt{-g}{\cal L}_m^{(2)}&=&{\kappa^2\over
2}\left(h^{(1)\mu\lambda}{h^{(1)\nu}}_\lambda -{1\over
2}h^{(1)}h^{(1)\mu\nu}\right)\partial_\mu\phi\partial_\nu\phi\nonumber\\
&-&{\kappa^2\over
8}\left(h^{(1)\alpha\beta}h^{(1)}_{\alpha\beta}-{1\over 2}
h^{(1)2}\right)(\partial^\alpha\phi\partial_\alpha\phi-m^2\phi^2)\label{eq:mlg}
\end{eqnarray}
The one- and two-graviton vertices are then respectively
\begin{eqnarray}
\tau_{\alpha\beta}(p,p')&=&{-i\kappa\over
2}\left(p_\alpha{p'}_\beta+{p'}_\alpha
p_\beta-\eta_{\alpha\beta}(p\cdot p'-m^2)\right)\nonumber\\
\tau_{\alpha\beta,\gamma\delta}(p,p')&=&i\kappa^2\left[
I_{\alpha\beta,\rho\xi}I^\xi{}_{\sigma,\gamma\delta}
\left(p^\rho{p'}^\sigma+{p'}^\rho p^\sigma\right)\right.\nonumber\\
&-&\left.{1\over
2}\left(\eta_{\alpha\beta}I_{\rho\sigma,\gamma\delta}
+\eta_{\gamma\delta}I_{\rho\sigma,\alpha\beta}\right){p'}^\rho
p^\sigma\right.\nonumber\\
&-&\left.{1\over 2}\left(I_{\alpha\beta,\gamma\delta} -{1\over
2}\eta_{\alpha\beta}\eta_{\gamma\delta}\right) \left(p\cdot
p'-m^2\right)\right]
\end{eqnarray}
where we have defined
$$I_{\alpha\beta;\gamma\delta}={1\over 2}(\eta_{\alpha\gamma}\eta_{\beta\delta}
+\eta_{\alpha\delta}\eta_{\beta\gamma})$$
We also require the triple graviton vertex
$\tau_{\alpha\beta,\gamma\delta}^{\mu\nu}(k,q)$ whose form is
\begin{eqnarray}
\tau^{\mu\nu}_{\alpha\beta,\gamma\delta}(k,q)&=&{i\kappa\over
2}\left\{ (I_{\alpha\beta,\gamma\delta}-{1\over
2}\eta_{\alpha\beta}\eta_{\gamma\delta})\left[k^\mu
k^\nu+(k-q)^\mu (k-q)^\nu+q^\mu q^\nu-{3\over
2}\eta^{\mu\nu}q^2\right]\right.\nonumber\\
&+&\left.2q_\lambda
q_\sigma\left[I^{\lambda\sigma,}{}_{\alpha\beta}I^{\mu\nu,}
{}_{\gamma\delta}+I^{\lambda\sigma,}{}_{\gamma\delta}I^{\mu\nu,}
{}_{\alpha\beta}-I^{\lambda\mu,}{}_{\alpha\beta}I^{\sigma\nu,}
{}_{\gamma\delta}-I^{\sigma\nu,}{}_{\alpha\beta}I^{\lambda\mu,}
{}_{\gamma\delta}\right]\right.\nonumber\\
&+&\left.[q_\lambda
q^\mu(\eta_{\alpha\beta}I^{\lambda\nu,}{}_{\gamma\delta}
+\eta_{\gamma\delta}I^{\lambda\nu,}{}_{\alpha\beta})+ q_\lambda
q^\nu(\eta_{\alpha\beta}I^{\lambda\mu,}{}_{\gamma\delta}
+\eta_{\gamma\delta}I^{\lambda\mu,}{}_{\alpha\beta})\right.\nonumber\\
&-&\left.q^2(\eta_{\alpha\beta}I^{\mu\nu,}{}_{\gamma\delta}+\eta_{\gamma\delta}
I^{\mu\nu,}{}_{\alpha\beta})-\eta^{\mu\nu}q^\lambda
q^\sigma(\eta_{\alpha\beta}
I_{\gamma\delta,\lambda\sigma}+\eta_{\gamma\delta}
I_{\alpha\beta,\lambda\sigma})]\right.\nonumber\\
&+&\left.[2q^\lambda(I^{\sigma\nu,}{}_{\alpha\beta}
I_{\gamma\delta,\lambda\sigma}(k-q)^\mu
+I^{\sigma\mu,}{}_{\alpha\beta}I_{\gamma\delta,\lambda\sigma}(k-q)^\nu\right.\nonumber\\
&-&\left.I^{\sigma\nu,}{}_{\gamma\delta}I_{\alpha\beta,\lambda\sigma}k^\mu-
I^{\sigma\mu,}{}_{\gamma\delta}I_{\alpha\beta,\lambda\sigma}k^\nu)\right.\nonumber\\
&+&\left.q^2(I^{\sigma\mu,}{}_{\alpha\beta}I_{\gamma\delta,\sigma}{}^\nu+
I_{\alpha\beta,\sigma}{}^\nu
I^{\sigma\mu,}{}_{\gamma\delta})+\eta^{\mu\nu}q^\lambda q_\sigma
(I_{\alpha\beta,\lambda\rho}I^{\rho\sigma,}{}_{\gamma\delta}+
I_{\gamma\delta,\lambda\rho}I^{\rho\sigma,}{}_{\alpha\beta})]\right.\nonumber\\
&+&\left.[(k^2+(k-q)^2)\left(I^{\sigma\mu,}{}_{\alpha\beta}I_{\gamma\delta,\sigma}{}^\nu
+I^{\sigma\nu,}{}_{\alpha\beta}I_{\gamma\delta,\sigma}{}^\mu-{1\over
2}\eta^{\mu\nu}P_{\alpha\beta,\gamma\delta}\right)\right.\nonumber\\
&-&\left.(k^2\eta_{\gamma\delta}I^{\mu\nu,}{}_{\alpha\beta}+(k-q)^2\eta_{\alpha\beta}
I^{\mu\nu,}{}_{\gamma\delta})]\right\}
\end{eqnarray}

\begin{center}
{\bf Gravitational Interactions: Spin 1/2}
\end{center}

 For the case of spin
1/2 we require some additional formalism in order to extract the
gravitational couplings.  In this case the matter Lagrangian reads
\begin{equation}
\sqrt{e}{\cal L}_m=\sqrt{e}\bar{\psi}(i\gamma^a{e_a}^\mu
D_\mu-m)\psi
\end{equation}
and involves the vierbein ${e_a}^\mu$ which links global
coordinates with those in a locally flat space\cite{ver1,ver2}.
The vierbein is in some sense the ``square root'' of the metric
tensor $g_{\mu\nu}$ and satisfies the relations
\begin{eqnarray}
{e^a}_\mu {e^b}_\nu\eta_{ab}&=&g_{\mu\nu},\qquad{e^a}_\mu
e_{a\nu}=g_{\mu\nu}\nonumber\\
e^{a\mu}e_{b\mu}&=&\delta^a_b,\qquad e^{a\mu}{e_a}^\nu=g^{\mu\nu}
\end{eqnarray}
The covariant derivative is defined via
\begin{equation}
D_\mu\psi=\partial_\mu\psi+{i\over 4}\sigma^{ab}\omega_{\mu ab}
\end{equation}
where
\begin{eqnarray}
\omega_{\mu ab}&=&{1\over 2}{e_a}^\nu(\partial_\mu
e_{b\nu}-\partial_\nu e_{b\mu})-{1\over 2}{e_b}^\nu(\partial_\mu
e_{a\nu}-\partial_\nu e_{a\mu})\nonumber\\
&+&{1\over 2}{e_a}^\rho{e_b}^\sigma(\partial_\sigma
e_{c\rho}-\partial_\rho e_{c\sigma}){e_\mu}^c
\end{eqnarray}
The connection with the metric tensor can be made via the
expansion
\begin{equation}
{e^a}_\mu=\delta^a_\mu+\kappa c^{(1)a}_\mu+\ldots
\end{equation}
The inverse of this matrix is
\begin{equation}
{e_a}^\mu=\delta_a^\mu-\kappa
c_a^{(1)\mu}+\kappa^2{c^{(1)\mu}_b{c^{(1)b}_a+\ldots}}
\end{equation}
and we find
\begin{equation}
g_{\mu\nu}=\eta_{\mu\nu}+\kappa c^{(1)}_{\mu\nu}+\kappa
c^{(1)}_{\nu\mu}+\ldots
\end{equation}
For our purposes we shall use only the symmetric component of the
c-matrices, since these are physical and can be connected to the
metric tensor.  We find then
$$c^{(1)}_{\mu\nu}\rightarrow {1\over
2}(c^{(1)}_{\mu\nu}+c^{(1)}_{\nu\mu})={1\over 2}h^{(1)}_{\mu\nu}$$
We have
\begin{equation}
{\rm det}\,e=1+\kappa c+\ldots\nonumber\\
=1+{\kappa\over 2}h+\ldots
\end{equation}
and, using these forms, the matter Lagrangian has the expansion
\begin{eqnarray}
\sqrt{e}{\cal L}_m^{(0)}&=&\bar{\psi}({i\over
2}\gamma^\alpha\delta^\mu_\alpha\overleftrightarrow{\nabla}_\mu-m)\psi\nonumber\\
\sqrt{e}{\cal L}_m^{(1)}&=&-{\kappa\over
2}h^{(1)\alpha\beta}\bar{\psi}i\gamma_\alpha
\overleftrightarrow{\nabla}_\beta\psi-{\kappa\over
2}h^{(1)}\bar{\psi}({i\over
2}\not\!{\overleftrightarrow{\nabla}}-m)\psi\nonumber\\
\sqrt{e}{\cal L}_m^{(2)}&=&{\kappa^2\over 8}h^{(1)}_{\alpha\beta}
h^{(1)\alpha\beta}\bar{\psi}i\gamma^\gamma\overleftrightarrow{
\nabla}_\lambda\psi+{\kappa^2\over
16}(h^{(1)})^2\bar{\psi}i\gamma^\gamma\overleftrightarrow{\nabla}_\gamma\psi\nonumber\\
&-&{\kappa^2\over
8}h^{(1)}\bar{\psi}i\gamma^\alpha{h_\alpha}^\lambda\overleftrightarrow{
\nabla}_\lambda\psi +{3\kappa^2\over
16}h_{\delta\alpha}^{(1)}h^{(1)\alpha\mu}\bar{\psi}i\gamma^\delta\overleftrightarrow{\nabla}_\mu\psi\nonumber\\
&+&{\kappa^2\over
4}h_{\alpha\beta}^{(1)}h^{(1)\alpha\beta}\bar{\psi}
m\psi-{\kappa^2\over
8}(h^{(1)})^2\bar{\psi}m\psi\nonumber\\
&+&{i\kappa^2\over 16}h_{\delta\nu}^{(1)}(\partial_\beta
h^{(1)\nu}_\alpha-\partial_\alpha h^{(1)\nu}_\beta)
\epsilon^{\alpha\beta\delta\epsilon}\bar{\psi}\gamma_\epsilon\gamma_5\psi
\end{eqnarray}
The corresponding one- and two-graviton vertices are found then to
be
\begin{eqnarray}
\tau_{\alpha\beta}(p,p')&=&{-i\kappa\over 2}\left[{1\over
4}(\gamma_\alpha(p+p')_\beta+\gamma_\beta(p+p')_\alpha)-{1\over
2}\eta_{\alpha\beta}({1\over
2}(\not\!\!{p}+\not\!\!{p}')-m)\right]\nonumber\\
\tau_{\alpha\beta,\gamma\delta}(p,p')&=&i\kappa^2\left\{-{1\over
2}({1\over
2}(\not\!\!{p}+\not\!\!{p}')-m)P_{\alpha\beta,\gamma\delta}\right.\nonumber\\
&-&\left.{1\over
16}[\eta_{\alpha\beta}(\gamma_\gamma(p+p')_\delta+\gamma_\delta(p+p')_\gamma)
\right.\nonumber\\
&+&\left.\eta_{\gamma\delta}(\gamma_\alpha
(p+p')_\beta+\gamma_\beta(p+p')_\alpha)]\right.\nonumber\\
&+&\left.{3\over
16}(p+p')^{\epsilon}\gamma^{\xi}(I_{\xi\phi,\alpha\beta}{I^{\phi}}_{\epsilon,\gamma\delta}
+I_{\xi\phi,\gamma\delta}{I^{\phi}}_{\epsilon,\alpha\beta})\right.\nonumber\\
&+&\left.{i\over 16}\epsilon^{\rho\sigma\eta\lambda}\gamma_\lambda
\gamma_5({I_{\alpha\beta,\eta}}^\nu
I_{\gamma\delta,\sigma\nu}{k'}_\rho-{I_{\gamma\delta,\eta}}^\nu
I_{\alpha\beta,\sigma\nu}k_\rho)\right\}
\end{eqnarray}

\begin{center}
{\bf Appendix B: Graviton Scattering Amplitudes}
\end{center}

In this section we summarize the independent contributions to the
various graviton scattering amplitudes which must be added in
order to produce the complete amplitudes quoted in the text.  We
leave it to the (perspicacious) reader to perform the appropriate
additions and to verify the factorized forms shown earlier.\\

\begin{center}
{\bf Graviton Photoproduction: Spin 0}
\end{center}

\begin{eqnarray}
{\rm Born-a}:&& {\rm Amp}_a=4e\kappa{(\epsilon_f^*\cdot
p_f)^2\epsilon_i\cdot
p_i\over 2p_i\cdot k_i}\nonumber\\
{\rm Born-b}:&& {\rm Amp}_b=-4e\kappa{(\epsilon_f^*\cdot
p_i)^2\epsilon_i\cdot
p_f\over 2p_i\cdot k_f}\nonumber\\
{\rm Seagull}:&&
{\rm Amp}_c=-2e\kappa\epsilon_f^*\cdot\epsilon_i\epsilon_f^*\cdot(p_i+p_f)\nonumber\\
{\rm \gamma-pole}:&&{\rm Amp}_d={e\kappa\over k_i\cdot
k_f}[\epsilon_f^*\cdot(p_i+p_f)(k_i\cdot
k_f\epsilon_f^*\cdot\epsilon_i-\epsilon_f^*\cdot
k_i\epsilon_i\cdot k_f)\nonumber\\
&+&\epsilon_f^*\cdot
k_i(\epsilon_f^*\cdot\epsilon_ik_i\cdot(p_i+p_f)-\epsilon_f^*\cdot
k_i\epsilon_i\cdot(p_i+p_f)]
\end{eqnarray}\\

\begin{center}
{\bf Graviton Photoproduction: Spin 1/2}
\end{center}

\begin{eqnarray}
{\rm Born-a}:&&{\rm Amp}_a=e\kappa{\epsilon_f^*\cdot p_f\over
2p_i\cdot
k_i}\bar{u}(p_f)[\not\!{\epsilon_f}^*(\not\!{p}_i+\not\!{k}_i+m)\not\!{\epsilon}_i]u(p_i)\nonumber\\
{\rm Born-b}:&&{\rm Amp}_b=-e\kappa{\epsilon_f^*\cdot p_i\over
2p_i\cdot
k_f}\bar{u}(p_f)[\not\!{\epsilon_i}(\not\!{p}_i-\not\!{k}_f+m)\not\!{\epsilon}_f^*]u(p_i)\nonumber\\
{\rm Seagull}:&&{\rm Amp}_c=-e\kappa\bar{u}(p_f)\not\!{\epsilon}_f^*u(p_i)\nonumber\\
{\rm \gamma-pole}:&&{\rm Amp}_d=e\kappa{1\over k_i\cdot
k_f}\bar{u}(p_f)[\not\!{\epsilon}_f^*(k_i\cdot
k_f\epsilon_f^*\cdot \epsilon_i-\epsilon_f^*\cdot
k_i\epsilon_i\cdot k_f)\nonumber\\
&+&\not\!{k}_f\epsilon_f^*\cdot\epsilon_i\epsilon_f^*\cdot
k_i-\not\!{\epsilon}_i(\epsilon_f^*\cdot k_i)^2]u(p_i)
\end{eqnarray}\\

\begin{center}
{\bf Graviton Scattering: Spin 0}
\end{center}

\begin{eqnarray}
{\rm Born-a}:&& {\rm Amp}_a=2\kappa^2{(\epsilon_i\cdot
p_i)^2(\epsilon_f^*\cdot
p_f)^2\over p_i\cdot k_i}\nonumber\\
{\rm Born-b}:&& {\rm Amp}_b=-2\kappa^2{(\epsilon_f^*\cdot
p_i)^2(\epsilon_i\cdot
p_f)^2\over p_i\cdot k_f}\nonumber\\
{\rm Seagull}:&&{\rm
Amp}_c=\kappa^2\left[\epsilon_f^*\cdot\epsilon_i(\epsilon_i\cdot
p_i\epsilon_f^*\cdot p_f+\epsilon_i\cdot p_f\epsilon_f^*\cdot
p_i)-{1\over 2}k_i\cdot
k_f(\epsilon_f^*\cdot\epsilon_i)^2\right]\nonumber\\
{\rm g-pole}:&&{\rm Amp}_d={4\kappa^2\over k_i\cdot k_f}
\left[\epsilon_f^*\cdot p_f\epsilon_f^*\cdot
p_i(\epsilon_i\cdot(p_i-p_f))^2+\epsilon_i\cdot p_i\epsilon_i\cdot
p_f(\epsilon_f^*\cdot(p_i+p_f))^2\right.\nonumber\\
&+&\left.\epsilon_i\cdot(p_i-p_f)\epsilon_f^*\cdot(p_f-p_i)(\epsilon_f^*\cdot
p_f\epsilon_i\cdot p_i+\epsilon_f^*\cdot p_i\epsilon_i\cdot
p_f)\right.\nonumber\\
&-&\left.\epsilon_f^*\cdot\epsilon_i\left(\epsilon_i\cdot(p_i-p_f)\epsilon_f^*\cdot
(p_f-p_i)(p_i\cdot p_f-m^2)\right.\right.\nonumber\\
&+&\left.\left.k_i\cdot k_f(\epsilon_f^*\cdot p_f\epsilon_i\cdot
p_i+\epsilon_f^*\cdot p_i\epsilon_i\cdot p_f)
+\epsilon_i\cdot(p_i-p_f)(\epsilon_f^*\cdot p_fp_i\cdot
k_f+\epsilon_f^*\cdot p_ip_f\cdot k_f)\right.\right.\nonumber\\
&+&\left.\left.\epsilon_f^*\cdot(p_f-p_i)(\epsilon_i\cdot
p_ip_f\cdot
k_i+\epsilon_i\cdot p_fp_i\cdot k_i)\right)\right.\nonumber\\
&+&\left.(\epsilon_f^*\cdot\epsilon_i)^2\left(p_i\cdot k_ip_f\cdot
k_i+p_i\cdot k_fp_f\cdot k_f-{1\over 2}(p_i\cdot k_ip_f\cdot
k_f+p_i\cdot k_fp_f\cdot k_i)\right.\right.\nonumber\\
&+&\left.\left.{3\over 2}k_i\cdot k_f(p_i\cdot
p_f-m^2)^2\right)\right]
\end{eqnarray}\\

\begin{center}
{\bf Graviton Scattering: Spin 1/2}
\end{center}

\begin{eqnarray}
{\rm Born-a}:&&{\rm Amp}_a=\kappa^2{\epsilon_f^*\cdot
p_f\epsilon_i\cdot p_i\over 8p_i\cdot
k_i}\bar{u}(p_f)[\not\!{\epsilon_f}^*(\not\!{p}_i+\not\!{k}_i+m)\not\!{\epsilon}_i]u(p_i)\nonumber\\
{\rm Born-b}:&&{\rm Amp}_b=-\kappa^2{\epsilon_f^*\cdot
p_i\epsilon_i\cdot p_f\over 8p_i\cdot
k_f}\bar{u}(p_f)[\not\!{\epsilon_i}(\not\!{p}_i-\not\!{k}_f+m)\not\!{\epsilon}_f^*]u(p_i)\nonumber\\
{\rm Seagull}:&& {\rm Amp}_c=\kappa^2\bar{u}(p_f)\left[{3\over
16}\epsilon_f^*\cdot\epsilon_i(\not\!{\epsilon_i}\epsilon_f^*\cdot(p_i+p_f)
+\not\!{\epsilon}_f^*\epsilon_i\cdot(p_i+p_f))\right.\nonumber\\
&+&\left.{i\over
8}\epsilon_f^*\cdot\epsilon_i\epsilon^{\rho\sigma\eta\lambda}
\gamma_\lambda\gamma_5(\epsilon_{i\eta}\epsilon_{f\sigma}^*k_{f\rho}-
\epsilon_{f\eta}^*\epsilon_{i\sigma}k_{i\rho})\right]u(p_i)\nonumber\\
{\rm g-pole}:&&{\rm Amp}_d={\kappa^2\over k_i\cdot
k_f}\bar{u}(p_f)\left[(\not\!{\epsilon}_i\epsilon_f^*\cdot
k_i+\not\!{\epsilon}_f^*\epsilon_i\cdot k_f)(\epsilon_i\cdot
p_i\epsilon_f^*\cdot p_f-\epsilon_i\cdot p_f\epsilon_f^*\cdot
p_i)\right.\nonumber\\
&-&\left.(\epsilon_f^*\cdot\epsilon_i)\left(k_i\cdot
k_f(\not\!{\epsilon}_f^*\epsilon_i\cdot
k_f+\not\!{\epsilon}_i\epsilon_f^*\cdot
p_i)\right.\right.\nonumber\\
&+&\left.\left.\not\!{k}_i(\epsilon_f^*\cdot p_f\epsilon_i\cdot
p_i-\epsilon_f^*\cdot p_i\epsilon_i\cdot p_f) +p_i\cdot
k_i(\not\!{\epsilon}_i\epsilon_f^*\cdot
k_i+\not\!{\epsilon}_f^*\epsilon_i\cdot
k_f)\right)\right.\nonumber\\
&+&\left.(\epsilon_f^*\cdot\epsilon_i)^2\not\!{k}_i(p_i\cdot
k_i-{1\over 2}k_i\cdot k_f)\right]u(p_i)
\end{eqnarray}\\

\begin{center}
{\bf Acknowledgement}
\end{center}

This work was supported in part by the National Science Foundation
under award PHY02-44801.  Thanks to Prof. A Faessler and the
theoretical physics group at the University of T\"{u}bingen, where
this work was completed, for hospitality.

\end{document}